\documentclass[11pt]{amsart}
\usepackage{amssymb}

\usepackage{amsmath,amssymb}
\usepackage{hyperref, color}

\numberwithin{equation}{section}

\newcommand{\R}{\mathbb R}
\begin{document}

\title{Energy, momentum, and center of mass in general relativity}
\author{Mu-Tao Wang}
\date{\today}
\begin{abstract}
 These notions in the title are of fundamental importance in any branch of physics. However, there have been great difficulties in finding physically acceptable definitions of them in general relativity since Einstein's time. I shall explain these difficulties and progresses that have been made. In particular, I shall introduce new definitions of center of mass and angular momentum at both the quasi-local and total levels, which are derived from first principles in general relativity and by the method of geometric analysis. With these new definitions, the classical formula $p=mv$  is shown to be consistent with Einstein's field equation for the first time. This paper is based on joint work \cite{Chen-Wang-Yau3, Chen-Wang-Yau4} with Po-Ning Chen and Shing-Tung Yau. 
\end{abstract}

\address{Department of Mathematics \\ Columbia University \\ 2990 Broadway \\ New York, NY 10027}

\thanks{This work was partially supported by a grant from the Simons Foundation (\#305519 to Mu-Tao Wang). The  author was partially supported by the National Science Foundation under grants DMS-1105483 and DMS-1405152.  He would like to thank his
collaborators Po-Ning Chen and Shing-Tung Yau, to whom he owes gratitude. 
}

\maketitle
\section{Introduction}

 Einstein's theory stipulates that the spacetime is a 4 dimensional Lorentzian manifold $(N^{3,1}, \bar{g})$, where $\bar{g}$ is a Lorentzian metric, or a symmetric 2-tensor of signature $(-, +, +, +)$. 
 For the purpose of exposition, we specialize to the vacuum case in this article and thus there is no presence of matter fields.   The spacetime metric $\bar{g}$ then satisfies the vacuum Einstein equation $Ric(\bar{g})=0$. The scenario is purely gravitational, or purely geometrical.
 The simplest example is  the Minkowski spacetime $\R^{3,1}$ where the metric is of the diagonal form $\bar{g}=(-1, 1,1,1)$. This is the  spacetime of special relativity where there is no gravitation and the curvature tensor vanishes. Other important examples include the Schwarzschild and the Kerr spacetimes. We refer to the survey article
\cite{MathGR} for basic materials in general relativity. 

In order to solve the Einstein equation, one considers the initial value formulation \cite{Choquet-Bruhat, Christodoulou-Klainerman} of the equation $Ric(\bar{g})=0$, and the spacetime is developed from an {\it initial data set}. 
Suppose the spacetime is foliated by the level sets of a time function $t$ and 
let $g_t$ be the 3-metric on each constant time slice, $Ric(\bar{g})=0$ is roughly 
\[\frac{\partial^2}{\partial t^2} g_t \sim \Delta_t g_t ,\] where $\Delta_t$ is the Laplace operator of $g_t$.  This is a second order nonlinear hyperbolic
system. Thus we need to prescribe $g_0$ on the initial time slice and the first time derivative of $g_0$,  which is essentially the second
fundamental form of the initial time slice, $k_0\sim \frac{\partial}{\partial t} g_t|_{t=0}$. This interpretation implies that there is a compatibility condition of hypersurfaces for the data $(g_0, k_0)$.   The condition is the so-called vacuum constrain equation 
for $(g, k)$: \begin{equation}\label{vc} \frac{1}{2}(R+(tr_g k)^2-|k|_g^2)=0 , div_g (k-(tr_g k)g) =0,\end{equation} where $R$ is the scalar curvature of $g$. Such an $(M, g, k)$ is called a {\it vacuum initial data set}. It can be shown that the vacuum constraint equation is satisfied for each subsequent time slice under suitable boundary conditions.

A natural boundary condition for the initial value problem of  the Einstein equation is ``asymptotically flatness". This corresponds to an isolated gravitating system around a planet, a star, or a black hole, which is far away from any other celestial bodies. We say an initial data set $(M, g, k)$ is {\it asymptotically flat}, if  there exists a compact subset $K$ of $M$ such that $M\backslash K$ is diffeomorphic to a finite union $\cup ( \R^3\backslash B)$ of ball complements of $\R^3$. Each complement $ \R^3\backslash B$ is an {\it end} and the
diffeomorphism on each end provides a coordinate system $(x^1, x^2, x^3)$, such that in this coordinate system
\begin{equation}\label{af} g_{ij}=\delta_{ij}+O(r^{-1}), k_{ij}=O(r^{-2}),\end{equation} where $r=\sqrt{\sum_{i=1}^3 (x^i)^2}$ and $\delta_{ij}$ is the Euclidean metric on $ \R^3\backslash B$. Suitable decay conditions for the derivatives of $g$ and $k$ need to be imposed as well, but we will be implicit about these conditions in this paper. 

From now on, we shall assume an initial data set $(M, g, k)$ is asymptotically flat and satisfies the vacuum constraint equation. Both conditions are 
preserved by the vacuum Einstein equation in the evolution \cite{Christodoulou}.

It is now natural to ask the total mass and energy for such a system $(M, g, k)$.  The ADM energy-momentum \cite{ADM} $(E, P_1, P_2, P_3)$ on each end is defined to be 
\begin{equation} \label{adm_E}E=\lim_{r\rightarrow \infty}\frac{1}{16\pi}
\int_{S_r}( g_{ij, j}-g_{jj,i})\nu^i\end{equation} and 
\begin{equation} \label{adm_P}P_j=\lim_{r\rightarrow \infty} \frac{1}{16\pi}\int_{S_r}
2(k_{ij}-\delta_{ij} tr_g k)\nu^i, j=1, 2, 3.\end{equation} Here $S_r$ is a coordinate sphere of radius $r$ with respect to the coordinate system on an end and $\nu^i$ is the unit outward normal of $S_r$.

 The most important property of the ADM energy-momentum is the positivity, which is now known as the positive mass theorem of Schoen-Yau \cite{Schoen-Yau} and Witten \cite{Witten}. A special case of the theorem implies that:

If $(M, g, k)$ is an asymptotically flat vacuum initial data set, then the ADM energy-momentum of each end satisfies  $E\geq 0$ and $E^2-\sum_{i=1}^3 (P_i)^2\geq 0$.  

 In particular, $m=\sqrt{E^2-\sum_{i=1}^3 (P_i)^2}$ is  the ADM mass of the system. The rigidity statement of the theorem asserts that $m=0$ if and only if  $(M, g, k)$ is a Minkowskian data.

Before proceeding further to ask for the center of mass of this system, one naturally wonders  why the mass expression is so different from the usual definition. Recall that in Newtonian gravity, the equation is $\Delta \Phi=4\pi \rho$, where $\rho$
is the mass density. For a domain $\Omega^3\subset \R^3$, the total mass is 
\begin{equation}\label{newtonian_mass} \int_\Omega \rho\end{equation} and the center of mass integral is 
\begin{equation} \label{newtonian_com} \int_\Omega  x^k  \rho, \end{equation} where $x^k, k=1, 2, 3$ are the standard coordinate functions on $\R^3$. 

However, in general relativity, a fundamental observation of Einstein is the so-called equivalence principle: THERE IS NO DENSITY FOR GRAVITATION. Therefore the formula that 
\[ \text{mass}= \int_\Omega \text{mass density} \] does NOT hold true for gravitation. Mathematically, this means at any point in spacetime, one can choose
a normal coordinate system so that all first derivatives of the Lorentzian metric $\bar{g}$ vanish at this point.

With this in mind, let us briefly review some existing definitions of center of mass in general relativity. The list works for exposition only and is not intended to be complete. 

 Huisken-Yau definition \cite{Huisken-Yau}: Suppose $(M, g)$ is complete Riemannian 3-manifold that is asymptotically Schwarzschildean in the sense  that the metric $g=(1+\frac{m}{2r}) \delta+O(r^{-2})$ and $m>0$ (again with suitable decay conditions on the derivatives of $g$), then there exists a unique CMC foliation by 2-surfaces $\Sigma_r$ of constant mean curvature in $M$ when $r$ is large enough. This foliation thus defines a family of canonical ``coordinate spheres" on each end.  
 The Huisken-Yau center of mass is defined to be 
\[\lim_{r\rightarrow \infty} \frac{\int_{\Sigma_r} x^k}{|\Sigma_r|}.\]

The definition pioneered a very active area of research with an extensive volume of literatures by now \cite{Ye, Rigger, Qing-Tian, Metzger, Neves-Tian, Huang, Eichmair-Metzger, Brendle-Eichmair, Cederbaum-Nerz, Nerz, Chan-Tam}.

Another definition is given by Regge-Teitelboim \cite{Regge-Teitelboim}  (see also \cite{Beig-Omurchadha}): For each end of a vacuum
asymptotically flat initial data set in the sense of \eqref{af}, the center of mass integral is defined to be:

\begin{align} \label{de:center-of-mass}
\begin{split} \lim_{r\rightarrow \infty}  \frac{1}{ 16 \pi } \int_{S_r}  x^k \left( g_{ij, j}- g_{jj,i} \right)\nu^i -(g_{ik} -\delta_{ik})\nu^i +(g_{ii} - \delta_{ii})\nu^k
\end{split}
\end{align}
We note that although the definition is intended for an initial data set $(M, g, k)$, only the metric $g$ appears in the expression.
This is clearly closely related to the ADM definition of mass. There are important gluing constructions to prescribe the center of mass integral by Corvino-Schoen \cite{Corvino-Schoen}, Chrusciel-Delay \cite{Chrusciel-Delay}, and Huang-Schoen-Wang \cite{Huang-Schoen-Wang}. 

The two definitions are shown to be equivalent in many cases \cite{Huang}. However, each one has its advantage and its problems. In particular, there are issues of  well-definedness and finiteness \cite{Cederbaum-Nerz, Nerz, Chan-Tam, Brendle-Eichmair}, in addition to the physical validity of the definition \cite{Chen-Huang-Wang-Yau}.  For example,  expression \eqref{de:center-of-mass} diverges apparently for standard asymptotics and a parity (Regge-Teitelboim) condition needs to be imposed to guarantee finiteness.  
More importantly, these are a priori abstract definitions and one naturally wonders the physical meaning of them. In particular, how are they related to the dynamics of the Einstein equation? 
At this point, let us discuss another definition pertaining to this issue.  
 In \cite{Christodoulou},  Christodoulou considered a ``strongly asymptotically flat" initial data set  $(M, g, k)$ on which $g=(1+\frac{m}{2r})\delta+O(r^{-1-\epsilon})$ and  $k=O(r^{-2-\epsilon})$ for a small positive constant $\epsilon$. The condition implies the ADM linear momentum \eqref{adm_P} $(P_1, P_2, P_3)=(0, 0, 0)$.  Under this condition,  it is shown that there exists a definition of  center of mass that is conserved along the Einstein equation. 

\section{The new definition}

 In joint work with Po-Ning  Chen and Shing-Tung Yau \cite{Chen-Wang-Yau3, Chen-Wang-Yau4}, we introduced new definitions of center of mass $C^i, i=1, 2, 3$ and angular momentum $J^i, i=1, 2, 3$ that satisfy rather remarkable properties. For example, we prove a finiteness theorem \cite[Section 7]{Chen-Wang-Yau3} under an order expansion condition, and in particular, no parity assumption is needed. More importantly, we show that 
 \[P_i= E \cdot \partial_t C^i\] when $(M, g(t), k(t))$ is a family of vacuum initial data sets evolving by the Einstein equation where $P_i$ is the ADM linear momentum and $E$ is the ADM energy.  This is the relativistic version of the classical formula $p=mv$. As fundamental as it is, this seems to be the first time when it is shown to be consistent with the nonlinear Einstein evolution. 

In addition, we prove an invariance of angular momentum theorem in the Kerr spacetime  \cite[Section 8]{Chen-Wang-Yau3}. It is likely that the finiteness theorem can be generalized to more general asymptotics (with the order expansion condition) using the gravitational conservation law in \cite[Section 5]{Chen-Wang-Yau3}.

 To motivate our definition, let's look at conserved quantities in the theory of special relativity for a moment. In special relativity,  a matter field in the Minkowksi spacetime $\R^{3,1}$ is equipped with the energy-momentum tensor of matter density $T$. Noether's principle asserts that each continuous symmetry of the theory corresponds to a conserved quantity. There are $10$ dimensional Killing fields on $\R^{3,1}$ that represent the infinitesimal isometry.  Each $K$ corresponds to a conserved quantity \[\int_{\Omega^3} T(K, u)\] where $u$ is unit normal of $\Omega^3$ for $\Omega^3\subset \R^{3,1}$. In the standard coordinate system $(t, x^1, x^2, x^3)$ on $\R^{3,1}$, the timelike translating Killing field $\frac{\partial}{\partial t}$ corresponds to the energy, so does any Killing field in the orbit of $\frac{\partial}{\partial t}$ under the isometric action of $SO(3,1)$. 
 
Similarly, each spacelike translating Killing in the orbit of $\frac{\partial}{\partial x^i}, i=1, 2, 3$ corresponds to a component of the linear momentum. Each rotation Killing field in the orbit of  $x^i\frac{\partial}{\partial x^j}-x^j\frac{\partial}{\partial x^i}, i<j$ corresponds to a component of the angular momentum. Each boost Killing field in the orbit of $x^i\frac{\partial}{\partial t}+t\frac{\partial}{\partial x^i}$ corresponds to a component of the center of mass.

However, in general relativity where the spacetime is in general curved, we encounter two major difficulties: there is  no density and there is (in a generic spacetime) no symmetry.

 The idea is to define conserved quantities quasi-locally on the boundary $\partial \Omega^3=\Sigma^2$ \cite{Penrose, Penrose2, Szabados}. We note that this is possible in Newtonian gravity. The total mass \eqref{newtonian_mass} equals to $\frac{1}{4\pi}\int_{\partial \Omega} \frac{\partial \Phi}{\partial \nu}$ by integrating by parts. This holds true for the center of mass integral \eqref{newtonian_com} as well. Once a quasi-local definition is available, we can take the limit along a foliation of surfaces to define total conserved quantities. On an asymptotically flat initial data set $(M, g, k)$, we look at conserve quantities on coordinates spheres $S_r$ and take the limit as $r\rightarrow \infty$. The same principle works for asymptotically hyperbolic and asymptotically null cases \cite{Chen-Wang-Yau1, Chen-Wang-Yau5}.

 To define conserved quantities quasi-locally, we start by looking for the ``best match" of the 2-surface of interest, $\Sigma \subset N$ (a physical spacetime) as a reference surface $\tilde{\Sigma}\subset \R^{3,1}$. A canonical identification of the normal bundles of the two surfaces is also needed to pull back symmetries (Killing fields) of $\R^{3,1}$ from $\tilde{\Sigma}\subset \R^{3,1}$ to $\Sigma\subset N$.

In the rest of this paper, we explain how to implement these ideas, which goes back to the proposal of optimal isometric embedding equations in \cite{Wang-Yau2}.

 Given a spacelike 2-surface $\Sigma$ with the topology of $S^2$ in a  spacetime $N$. Let $\sigma$ be the induced metric on $\Sigma$ and $H$ be the mean curvature vector field, which is the unique normal vector field on $\Sigma$ such that the variation of the area function $|\Sigma|$ satisfies
 \[\delta_V|\Sigma|=-\int_\Sigma\langle H, V\rangle\] for any variation field $V$. We assume $H$ is spacelike and extract from it a function $|H|$ and a connection one-form $\alpha_H$ of the normal bundle in the mean 
curvature gauge (see \cite{Chen-Wang-Yau3} for the definition of $\alpha_H$). $(\sigma, |H|, \alpha_H)$ determines the ``best match" surface $\tilde{\Sigma}\subset \R^{3,1}$. 

 Let $X:\Sigma\rightarrow \R^{3,1}$ be an isometric embedding of $\sigma$, i.e $\langle dX, dX\rangle =\sigma$. This is an under-determined PDE system with 4 unknowns (the coordinate functions of $\R^{3,1}$ ) and 3 equations (the components of $\sigma$). We need to impose one more equation in order to make the system well-determined.  This is the {\it optimal isometric embedding} 
 equation, which is obtained by a variational method. 
 
 Let $\tau$ be the time-component of $X$. $\tau=0$ identically corresponds to an isometric embedding into a totally geodesic $\R^3\subset \R^{3,1}$. This case was solved by the work of Nirenberg \cite{ni} and Pogorelov \cite{po}, and applied to define and study earlier proposals of quasi-local mass \cite{Brown-York2, Shi-Tam, Liu-Yau, ki}.  The optimal isometric embedding  essentially imposes an equation on $\tau$ in the general case in order to resolve some unwanted
difficulty of previous proposals (see \cite{Wang-Yau2}). Let $H_0$ be the mean curvature vector of the image surface $X(\Sigma)=\tilde{\Sigma}$ and again
we extract a function $|H_0|$ and a one-form $\alpha_{H_0}$.  Consider the following function $\rho$ and 1-form $j_a$ on $\Sigma$:
  \begin{equation} \label{rho} \begin{split}\rho &= \frac{\sqrt{|H_0|^2 +\frac{(\Delta \tau)^2}{1+ |\nabla \tau|^2}} - \sqrt{|H|^2 +\frac{(\Delta \tau)^2}{1+ |\nabla \tau|^2}} }{ \sqrt{1+ |\nabla \tau|^2}} \end{split}\end{equation} and
  \begin{equation} \label{j_a}
j_a=\rho {\nabla_a \tau }- \nabla_a [ \sinh^{-1} (\frac{\rho\Delta \tau }{|H_0||H|})]-(\alpha_{H_0})_a + (\alpha_{H})_a. \end{equation} 
Here $\Delta$ is the Laplace operator of the metric $\sigma$ and $a=1, 2$ denotes a coordinate index on $\Sigma$. 
 
 The quasi-local energy with respect to the isometric embedding $X$ with time component $\tau$ is \begin{equation}\label{qle} E(\Sigma, \tau)=\frac{1}{8\pi}\int_\Sigma (\rho+j_a\nabla^a\tau).\end{equation} Minimizing among possible $\tau$'s, the Euler-Lagrangian equation for critical points of $E$ is \begin{equation}\label{opt}\nabla^a j_a=0,\end{equation} see \cite{Chen-Wang-Yau3} for the derivation.  \eqref{opt} is exactly the additional optimal embedding equation imposed on an isometric embedding into $\R^{3,1}$. The expression \eqref{qle} originates from the boundary term of Hamilton-Jacobi analysis of the gravitational action \cite{Brown-York2, hh}. We refer to \cite{icmp} for more detailed discussions on this aspect. 

 For a critical point $X$ that satisfies the optimal isometric embedding equation, by integration by parts, the quasi-local energy becomes \[\frac{1}{8\pi}\int_\Sigma \rho.\]  $\rho$ is thus called the quasi-local mass density. The energy \eqref{qle} satisfies the important positivity and rigidity properties by the work of \cite{Wang-Yau1, Wang-Yau2}, see also \cite{Shi-Tam, Liu-Yau}.

 For each optimal isometric embedding $X:\Sigma\rightarrow \R^{3,1}$, i.e. $\langle dX, dX\rangle=\sigma$ and  $\nabla_\sigma j=0$, we obtain a pair $\rho$, $j_a$ that correspond to quasi-local energy-momentum density and a reference surface $X(\Sigma)=\tilde{\Sigma}$ in $\R^{3, 1}$ that is isometric to $\Sigma$.  We can thus proceed to define 
 quasi-local conserved quantities with respect to $X$.

 Given any Killing field $K$ in $\R^{3,1}$, we define the corresponding quasi-local conserved quantity to be: 

\[-\frac{1}{8\pi} \int_\Sigma
( \langle K, T_0\rangle \rho+(K^\top)^a  j_a),\] where $T_0=(1, 0, 0, 0)$ and $K^\top$ corresponds to the tangential part of $K$ to $\Sigma$. Strictly speaking the equasi-local energy \eqref{qle} depends on the pair $(X, T_0)$ where $T_0$ is a general future timelike translating Killing field in $\R^{3,1}$ and $\tau=-\langle X, T_0\rangle$ is the time component with respect to $T_0$. In this article, we only consider the case when $T_0=(1, 0, 0, 0)$ for simplicity. In general, we need to take into account of the Lorentz group action on Killing fields as in \cite{Chen-Wang-Yau3}.

 Therefore, a boost Killing field defines a component of the center of mass and a rotation Killing field defines a component of the angular momentum.  

For a given asymptotically flat vacuum initial data set $(M, g, k)$, we consider the quasi-local conserved quantities on coordinate sphere $S_r$. 
It is shown \cite{mt, Chen-Wang-Yau1, Chen-Wang-Yau2} that if the ADM mass of $(M, g, k)$ is positive, there is a unique, energy-minimizing, optimal isometric embedding of $S_r$ whose image approaches a large round sphere in $\R^3$. Take the limit as $r\rightarrow \infty$ of the quasi-local conserved quantities on $S_r$, we obtain $(E, P_i, J^i, C^i)$ where $(E, P_i)$ are the same as the ADM energy-momentum vector \cite{Wang-Yau3}. In comparison to the Huisken-Yau and Regge-Teitelboim definitions,
the new definition does involve the second fundamental form $k$ of an initial data set $(M, g, k)$, in which the dynamical information is encoded.

 When $(M, g(t), k(t))$ is evolved by the vacuum Einstein equation, we derived that $\partial_t J^i=0$ and $\partial_t C^i=\frac{P_i}{E}$ \cite{Chen-Wang-Yau3, Chen-Wang-Yau4}. The same result holds in  the presence of matter fields, as long as one assumed suitable decay conditions of the energy
momentum tensor at infinity, see also \cite{Nerz}. The proof is based on several ingredients:
 
1. The Einstein evolution equation and the vacuum constraint equation \eqref{vc}. 

2. The optimal isometric embedding equation \eqref{opt}. 

3. Conservation law of the quasi-local conserved quantity \cite[Section 5]{Chen-Wang-Yau3}.

We refer to the articles \cite{Chen-Wang-Yau3, Chen-Wang-Yau4}, in particular \cite[Section 9]{Chen-Wang-Yau3}, for more details.


\end{document}